\documentclass[]{JHEP3}   
 \preprint{  }

\usepackage{epsfig,multicol}
\usepackage{amsmath}
\usepackage{graphics}
\usepackage{graphicx}
\usepackage{dcolumn}
\usepackage{amssymb}
\usepackage{appendix}
\usepackage{slashbox}

\title{{\bf Probing spacetime noncommutative constant via charged astrophysical black hole lensing}}

\author{Chikun Ding \thanks{Email: dingchikun@163.com, QQ: 649433259}\\
 Department of Physics and Information Engineering, \\
Hunan Institute of Humanities, Science and Technology,\\ Loudi,
Hunan 417000, P. R. China}

\author{ Jiliang  Jing\thanks{Email:
jljing@hunnu.edu.cn}\\
 Department of Physics, and Key Laboratory of Low Dimensional Quantum Structures \\
and Quantum Control of Ministry of Education, Hunan Normal University,
\\ Changsha, Hunan 410081, P. R. China}

\abstract {

 We study the influence of the spacetime noncommutative
parameter on the strong field gravitational lensing in the
noncommutative Reissner-Nordstr\"{o}m black-hole spacetime.
Supposing that the gravitational field of the supermassive central
object of the Galaxy is described by this metric, we estimate the
numerical values of the coefficients and observables for strong
gravitational lensing. Our results show that with the increase of
the parameter $\sqrt{\vartheta}$, the observables $\theta_{\infty}$
and $r_m$ decrease, while $s$ increases. Our results also show that
i) if $\sqrt{\vartheta}$ is strong, the observables are close to
those of the noncommutative Schwarzschild black hole lensing; ii) if
$\sqrt{\vartheta}$ is weak, the observables are close to those of
the commutative Reissner-Nordstr\"{o}m black hole lensing; iii) the
detectable scope of $\vartheta$ in a noncommutative
Reissner-Nordstr\"{o}m black hole lensing is
$0.12\leq\sqrt{\vartheta}\leq0.26$, which is wider than that in a
noncommutative Schwarzschild black hole lensing,
$0.18\leq\sqrt{\vartheta}\leq0.26$. This may offer a way to probe
the spacetime noncommutative constant $\vartheta$ by the
astronomical instruments in the future. }


\begin{document}

\section{Introduction}

Quantum mechanics teach us that the emergence of a minimal length is
a natural requirement when quantum features of phase space are
considered. It also holds true to spacetime \cite{garattini}. The
presence of a minimal length implies that singularities in general
relativity and ultraviolet divergences in quantum field theory are
nothing but spurious effects due to the inadequacy of the formalism
at small scales/extreme energies, rather than actual physical
phenomena. Given this background, noncommutative geometry have been
devoted to implementing a minimal length in physical theories and
curing the aforementioned pathologies. In the string theory,
coordinates of the target spacetime become {\it noncommutating}
operators on a $D$-brane as \cite{seigerg}
\begin{equation}[\hat{x}^\mu,\hat{x}^\nu]=i\vartheta^{\mu\nu},\end{equation} where
$\vartheta^{\mu\nu}$  is a real, anti-symmetric and constant tensor
which determines the fundamental cell discretization of spacetime
much in the same way as the Planck constant $\hbar$ discretizes the
phase space, $[\hat{x}_i,\hat{p}_j]=i\hbar\delta_{ij}$. Motivated by
string theory arguments, noncommutative spacetime has been
reconsidered again and is believed to afford a starting point to
quantum gravity.

Noncommutative spacetime is not a new conception, and coordinate
noncommutativity also appears in another fields, such as in quantum
Hall effect \cite{bell}, the noncommutative Landau problem
\cite{gamboa}, cosmology \cite{marcoll}, the model of a very slowly
moving charged particle on a constant magnetic field \cite{kim}, a
strong magnetic field \cite{jackiw}, the Chern-Simon's theory
\cite{deser}, and so on.
 The idea of
noncommutative spacetime dates back to Snyder \cite{snyder} who used
the noncommutative structure of spacetime to introduce a small
length scale cut-off in field theory without breaking Lorentz
invariance and Yang \cite{yang} who extended Snyder's work to
quantize spacetime in 1947 before the renormalization theory.
Noncommutative geometry \cite{connes} is a branch of mathematics
that has many applications in physics, a good review of the
noncommutative spacetime is in \cite{connes2, akofor}.

The fundamental notion of the noncommutative geometry is that the
picture of spacetime as a manifold of points breaks down at distance
scales of the order of the Planck length: Spacetime events cannot be
localized with an accuracy given by Planck length \cite{akofor} as
well as particles do in the quantum phase space. So that the points
on the classical commutative manifold should then be replaced by
states on a noncommutative algebra and the point-like object is
replaced by a smeared object \cite{smail} to cure the singularity
problems at the terminal stage of black hole evaporation
\cite{nico}.

The approach to noncommutative quantum field theory follows two
paths: one is based on the Weyl-Wigner- Moyal *-product and the
other on coordinate coherent state formalism \cite{smail}. In a
recent paper \cite{morita}, following the coherent state approach,
it has been shown that Lorentz invariance and unitary, which are
controversial questions raised in the *-product approach, can be
achieved by assuming\begin{equation}\vartheta^{\mu\nu}=\vartheta \;
\text{diag}(\epsilon_1,\ldots, \epsilon_{D/2}),\end{equation} where
$\vartheta$ \cite{foot} is a constant which has the dimension of
$length^2$, $D$ is the dimension of spacetime \cite{smail2} and,
there isn't any UV/IR mixing. Inspire by these results, various
black hole solutions of noncommutative spacetime have been found
\cite{Nicolini:2008aj}; thermodynamic properties of the
noncommutative black hole were studied in \cite{nozari}; the
evaporation of the noncommutative black hole was studied in
\cite{comp}; quantum tunneling of noncommutative Kerr black hole was
studied in \cite{miao}; quantized entropy was studied in \cite{wei},
and so on.

Noncommutative black holes are currently the richest class of
quantum gravity black holes \cite{mann} which are connected with a
recently proposed ultraviolet complete quantum gravity \cite{moffat}
and, have been recently taken into account in Monte Carlo
simulations as reliable candidate models to describe the conjectured
production of microscopic black holes in particle accelerators
\cite{gingrich}. It is interesting that the noncommutative spacetime
coordinates introduce a new fundamental natural length scale
\begin{equation}l_{NC}=\sqrt{\vartheta},\end{equation}
the effects of noncommutativity are increasing important from
main-sequence stars to neutron stars and it might have a relevant
impact on black-hole physics \cite{berto}. In this paper we will
continue to study its influence on strong gravitational lensing.

In previous work \cite{ding}, we supposed that the gravitational
field of the supermassive central object of the Galaxy can be
described by the noncommutative Schwarzschild metric, and estimated
the numerical values of the coefficients and observables for strong
gravitational lensing. In comparison to the commutative
Reissner-Nordstr\"{o}m black hole, we find that the influences of
the spacetime noncommutative parameter is similar to those of the
charge, but these influences are much smaller. However some problems
would appear: (i) if the spacetime noncommutative constant
$\vartheta$ is very strong or the charge quantity of the black hole
is very small, it is difficult to distinguish the influences of the
$\vartheta$ from the charge; (ii) if the spacetime noncommutative
constant $\vartheta$ is very weak, the study of the noncommutative
Schwarzschild lensing cannot probe it at all since that its effects
cannot obviously deviate from those of the commutative Schwarzschild
lensing until $\sqrt{\vartheta}\geq0.18$. Therefore it is not enough
to study the lensing in the noncommutative Schwarzschild black-hole
spacetime. Hence in this paper, we plan to study the influence of
this constant on strong gravitational lensing in the noncommutative
Reissner-Nordstr\"{o}m black-hole spacetime.

Gravitational lensing continues to be a major source of insight into
gravitation and cosmology \cite{wambsganss}. When the lens is a
black hole, a strong field treatment of gravitational lensing
\cite{Darwin,Vir,Vir1,Vir2,Vir3,Fritt} is needed. The relativistic
images of strong gravitational lensing could provide a profound
verification of alternative theories of gravity. Thus, the study of
the strong gravitational lensing becomes appealing recent years.
Bhadra \textit{et al} \cite{Bhad1}\cite{Sarkar} have considered the
Gibbons-Maeda-Garfinkle-Horowitz-Strominger black hole lensing.
Eiroa \textit{et al} \cite{Eirc1} have studied the
Reissner-Nordstr\"{o}m black hole lensing. Konoplya \cite{Konoplya1}
has studied the corrections to the deflection angle and time delay
of black hole lensing immersed in a uniform magnetic field. Majumdar
\cite{Muk} has investigated the dilaton-de Sitter black hole
lensing. Perlick \cite{Per} has obtained an exact lens equation and
used it to study
 Barriola-Vilenkin monopole black hole lensing.
 Bin-Nun
\cite{bin} studied Sagittarius A* (Sgr A*) lensing; Wei {\it et al}
studied the strong gravitational lensing in Kerr-Taub-NUT spacetime
\cite{shaowen}, braneworld black holes lensing were studied in
\cite{majumd}, and so on.

The plan of our paper is organized as follows. In Sec. II we study
some properties of the noncommutative Reissner-Nordstr\"{o}m black
hole metric. In Sec. III we adopt Bozza's method and obtain the
deflection angles for light rays propagating in the noncommutative
Reissner-Nordstr\"{o}m black hole spacetime. In Sec. IV we suppose
that the gravitational field of the supermassive black hole at the
centre of our Galaxy can be described by this metric and then obtain
the numerical results for the observational gravitational lensing
parameters defined in Sec. III. Then, we make a comparison among the
properties of gravitational lensing in the
noncommutative/commutative Reissner-Nordstr\"{o}m and noncommutative
Schwarzschild metrics. In Sec. V, we present a summary.

\section{The charged astrophysical black hole}

The Reissner-Nordstr\"{o}m solution is a natural result of the
Einstein-Maxwell field equation which shows that a black hole can
possess some net charge. But there exists a general consensus that
astrophysical objects with large amounts of charge can not exist in
nature since that astrophysical objects are always surrounded by
some plasma, which is a very good conductor \cite{eddington}. The
same reasoning is applied to black holes \cite{eddington} which
shows that the real astrophysical black holes should be neutral.
This point of view had been challenged by several researchers
\cite{bally}. Recently, Some authors found that the charged black
hole can be formed during charged stellar collapse \cite{ghezzi} or
accreting process of a neutral black bole \cite{diego}.

Using the brane-world-inspired charge-leaking mechanism, Cuesta {\it
et al.} \cite{cuesta} provide a natural explanation for the
formation of charged black holes after supernova collapse of massive
stars. In brane world models free particles localized on the brane
can leak out to the extra space. If there were color confinement in
the bulk, electrons would be more able to escape than quarks and
protons which generates an electric charge asymmetry on the brane
matter densities. A tiny charge asymmetry would generate small
electromagnetic fields which can be substantially magnified
\cite{shap} by some astrophysical processes, such as a supernova
core collapse. According to Ref. \cite{ruff}, vacuum polarization
occurring during the formation of a Reissner-Nordstr\"{o}m black
hole may generate gamma-ray bursts, provided that the net
charge-to-mass ratio, $\xi=Q/(2M\sqrt{G})$, be of the order 0.005 to
0.5.

   Therefore considering strong gravitational
lensing in charged black hole spacetime is practical. The line
element of the noncommutative Reissner-Nordstr\"{o}m black hole
reads \cite{nico}
\begin{eqnarray}
ds^2 = -f(r)\,dt^2 + \frac{dr^2}{f(r)} + r^2 (d\theta^2
+\sin^2\theta d\phi^2)\,, \label{metric}
\end{eqnarray}
and
\begin{eqnarray}
\label{sol1} f(r)&=&1- \frac{4GM}{c^2r\sqrt{\pi}}\,
\gamma(\frac{3}{2} \ , \frac{r^2}{4\vartheta}\,)+\frac{GQ^2}{\pi c^4
r^2}F(r),\,\nonumber\\
F(r)&=&\left[\gamma^2(\frac{1}{2} \ ,
\frac{r^2}{4\vartheta}\,)-\frac{r}{\sqrt{2\vartheta}}\gamma(\frac{1}{2}
\ ,
\frac{r^2}{2\vartheta}\,)+\sqrt{\frac{2}{\vartheta}}r\gamma(\frac{3}{2}
\ , \frac{r^2}{4\vartheta}\,)\right],
\end{eqnarray}
where $\gamma\left(s \ ,x\, \right)$ is the lower incomplete Gamma
function:
\begin{equation}
\gamma\left(s\ , x\, \right)\equiv \int_0^{x} dt\, t^{s-1} e^{-t},
\end{equation}
and $\vartheta$ is a spacetime noncommutative parameter \cite{foot},
$G$ is Newtonian constant, $c$ is light velocity in vacuum. The
commutative Reissner-Nordstr\"{o}m metric is obtained from
(\ref{metric}) in the limit $r/\sqrt{\vartheta}\to\infty $. Equation
(\ref{metric}) leads to the mass distribution $m\left(\, r\,\right)=
2M \,\gamma\left(3/2\ , r^2/4\vartheta\, \right)/\sqrt\pi $ and the
charge distribution $q\left(\, r\,\right)= Q
 \sqrt{F(r)/\pi} $, where $M$ is the total mass of the source and $Q$ is the total charge of the source.

Depending on the values of $Q$, $\sqrt{\vartheta}$ and $M$, the
metric displays different causal structure: existence of two
horizons (non-extremal black hole), one horizon (extremal black
hole) or no horizons (massive charged ¡°droplet¡± ). Due to
$f(r_+)=0$ cannot be solved analytically, we list some values of the
maximum charge $Q_{max}$ and the single horizon $r_+$ in Table
\ref{tabel1} by letting $2M=G=c=1$\footnote{The units we used here
and hereafter is the total mass of the black hole $2M$, i.e.,
$\frac{c^2r}{2GM}\rightarrow r, \;\frac{Q}{2M\sqrt{G}}\rightarrow
Q,\; \frac{c^2\sqrt{\vartheta}}{2GM}\rightarrow\sqrt{\vartheta}.$}.
\begin{table}[h] \caption{Numerical values for the
radius of the single event horizon in the charged noncommutative
black hole spacetime with different $\sqrt{\vartheta}$ and
$Q_{max}$. }\label{tabel1}
\begin{center}
\begin{tabular}{|c|c|c|c|c|c|c|c|}
\hline \hline $\sqrt{\vartheta}$ &0.262475&0.26 & 0.24 &0.22&
0.20&0.18&0.16 \\
\hline
 $Q_{max}$& 0& 0.09441&0.26858&0.35411& 0.41081& 0.44988&0.47593\\
\hline
 $r_{+}$&0.80802& 0.78803&0.74050&0.69423& 0.65005& 0.60673&0.56632\\
 \hline $\sqrt{\vartheta}$ &0.14&0.12&0.10& 0.08 &0.06&
0.04&0.02 \\
\hline
 $Q_{max}$& 0.49151& 0.49867&0.50000&0.50000& 0.50000& 0.50000&0.50000\\
\hline
 $r_{+}$& 0.53252&0.50794&0.51189&0.50509& 0.50022& 0.50000&0.50000\\
\hline\hline
\end{tabular}
\end{center}
\end{table}
Table \ref{tabel1} shows that the maximum charge $Q_{max}$ decreases
with the increase of the spacetime noncommutative parameter
$\sqrt{\vartheta}$. It indicates the restriction of the spacetime
non-commutativity on the charge of black hole which implies that i)
if $\sqrt{\vartheta}$ is strong, its single horizon is close to that
of the noncommutative Schwarzschild black hole; ii) if
$\sqrt{\vartheta}$ is weak, its single horizon is close to that of
the commutative Reissner-Nordstr\"{o}m black hole. When
$M>1.9\sqrt{\vartheta}$ and $0\leq Q<Q_{max}$, the two horizons
(non-extremal black hole) are given by
\begin{eqnarray}
r_\pm=\frac{2}{\sqrt{\pi}}\,\gamma\left(3/2\ , r^2_\pm/4\vartheta\,
\right)+\frac{Q^2}{\pi r_\pm}F(r_\pm).
\end{eqnarray}
which is different from the commutative Reissner-Nordstr\"{o}m black
hole. The line element (\ref{metric}) describes the geometry of a
noncommutative black hole and should give us useful insights about
possible spacetime noncommutative effects on strong gravitational
lensing.

\section{Deflection angle in the charged noncommutative black hole spacetime}
As in \cite{Vir2,Vir3,Bozza2}, the deflection angle for the photon
coming from infinite can be expressed as
\begin{eqnarray}
\alpha(r_0)=I(r_0)-\pi,
\end{eqnarray}
where $r_0$ is the closest approach distance and $I(r_0)$ is
\cite{Vir2,Vir3}
\begin{eqnarray}
I(r_0)=2\int^{\infty}_{r_0}\frac{\sqrt{B(r)}dr}{\sqrt{C(r)}
\sqrt{\frac{C(r)A(r_0)}{C(r_0)A(r)}-1}},\label{int1}
\end{eqnarray}with
\begin{eqnarray}
A(r)=f(r), \;\;\;\;B(r)&=&1/f(r),\;\;\;\; C(r)=r^2.
\end{eqnarray}

It is easy to determine that as parameter $r_0$ decreases, the
deflection angle increases. At a certain point, the deflection angle
will become $2\pi$; this means that the light ray will make a
complete loop around the compact object before reaching the
observer. When $r_0$ is equal to the radius of the photon sphere,
the deflection angle diverges and the photon is captured.

The photon sphere equation is given by \cite{Vir2,Vir3}
\begin{eqnarray}
\frac{C'(r)}{C(r)}=\frac{A'(r)}{A(r)},\label{root}
\end{eqnarray}
which admits at least one positive solution, and then the largest
real root of Eq. (\ref{root}) is defined  as the radius of the
photon sphere. To the noncommutative Reissner-Nordstr\"{o}m black
hole metric (\ref{metric}), the radius of the photon sphere can be
given implicitly by
\begin{eqnarray}\label{rpseq}
&&r_{ps}=\frac{3}{\sqrt{\pi}}\gamma(\frac{3}{2},
\frac{r_{ps}^2}{4\vartheta})-\frac{r_{ps}^3}{4\vartheta\sqrt{\pi\vartheta}}
e^{-\frac{r_{ps}^2}{4\vartheta}}-\frac{2Q^2}{\pi
r_{ps}}F(r_{ps})+\frac{Q^2}{\pi \sqrt{\vartheta}}G(r_{ps}),
\nonumber\\ &&G(r_{ps})=\left[\gamma(\frac{1}{2} \ ,
\frac{r_{ps}^2}{4\vartheta}\,)-\frac{r_{ps}}{2\sqrt{\vartheta}}e^{-\frac{r_{ps}^2}{4\vartheta}}
+\frac{r^3_{ps}}{4\vartheta\sqrt{2\vartheta}}\right]e^{-\frac{r_{ps}^2}{4\vartheta}}
-\frac{1}{\sqrt{2}}\left[\frac{1}{2}\gamma(\frac{1}{2} \
,\frac{r_{ps}^2}{2\vartheta}\,)-\gamma(\frac{3}{2} \ ,
\frac{r_{ps}^2}{4\vartheta}\,)\right],
\end{eqnarray}
which is an implicit function $f(r_{ps},Q,\sqrt{\vartheta})=0$. It
cannot be expressed as an explicit function
$r_{ps}=g(Q,\sqrt{\vartheta})$, so we list some values of the photon
sphere radius in the following table.
\begin{table}[h]
\caption{Numerical values for the radius of the photon sphere in the
charged noncommutative black hole spacetime with different
$\sqrt{\vartheta}$ and $Q$. For a certain $\sqrt{\vartheta}$ the
charge should satisfy the condition $0\leq Q<Q_{max}$, so that
corresponding slots are left vacant.}\label{tabel2}
\begin{center}
\begin{tabular}{|l|*{7}{c|}}\hline\backslashbox{$\sqrt{\vartheta}$}{$Q$}
&\makebox[3em]{0.0}&\makebox[3em]{0.09}&\makebox[3em]{0.26}
&\makebox[3em]{0.35}&\makebox[3em]{0.4}&\makebox[3em]{0.45}&\makebox[3em]{0.49}
 \\  \hline 0.0 &1.50000&1.48912&1.40368&1.31347& 1.24244&1.14686&1.03688\\
\hline 0.14 &1.50000&1.48912&1.40368&1.31347&1.24244&1.14686&1.03684\\
 \hline 0.16 &1.50000&1.48912&1.40368&1.31347&1.24243&1.14676&---\\
 \hline 0.18 &1.50000&1.48912&1.40367&1.31339&1.24216&1.14557&---\\
 \hline 0.20&1.49995&1.48906&1.40345&1.31263&1.24024&---&---\\
 \hline 0.22 &1.49954&1.48859&1.40215&1.30898&---&---&---\\
 \hline 0.24 &1.49764&1.48647&1.39714&---&---&---&---\\
 \hline 0.26 &1.49151&1.47971&---&---&---&---&---\\
\hline
\end{tabular}
\end{center}
\end{table}

From Table \ref{tabel2} we can see that, when
$\sqrt{\vartheta}\rightarrow0$, the photon sphere radius can recover
that in the commutative Reissner-Nordstr\"{o}m black hole spacetime,
i.e. $r_{ps}=(3+\sqrt{9-32Q^2})/4$. The presence of
$\sqrt{\vartheta}$ decreases the photon sphere radius as well as the
charge does. If $\sqrt{\vartheta}$ is weak, the photon sphere radius
is close to
 that of the commutative Reissner-Nordstr\"{o}m black-hole lens;
  if $\sqrt{\vartheta}$ is strong, it is close to
 that of the noncommutative Schwarzschild black-hole lens.
 All these features imply that
there exist some distinct effects of the noncommutative parameter
$\vartheta$ on gravitational lensing in the strong field limit.

\begin{figure}[ht]
\begin{center}
\includegraphics[width=7.5cm]{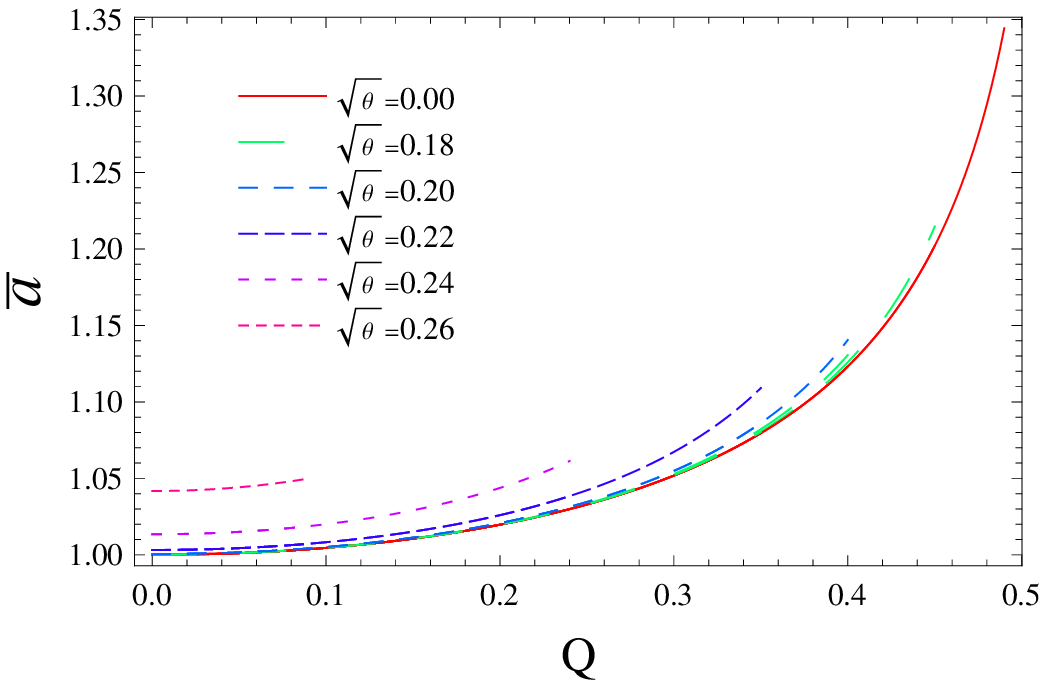}\;\;\;\;\; \includegraphics[width=7.5cm]{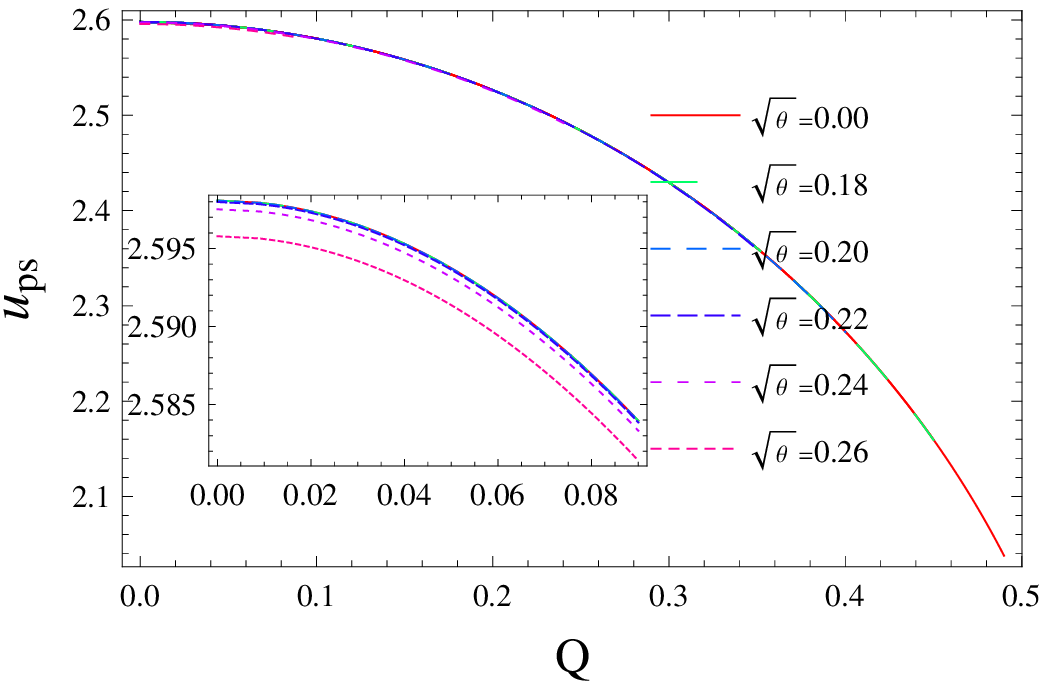}\;\;\;\;\;\\
 \includegraphics[width=11.0cm]{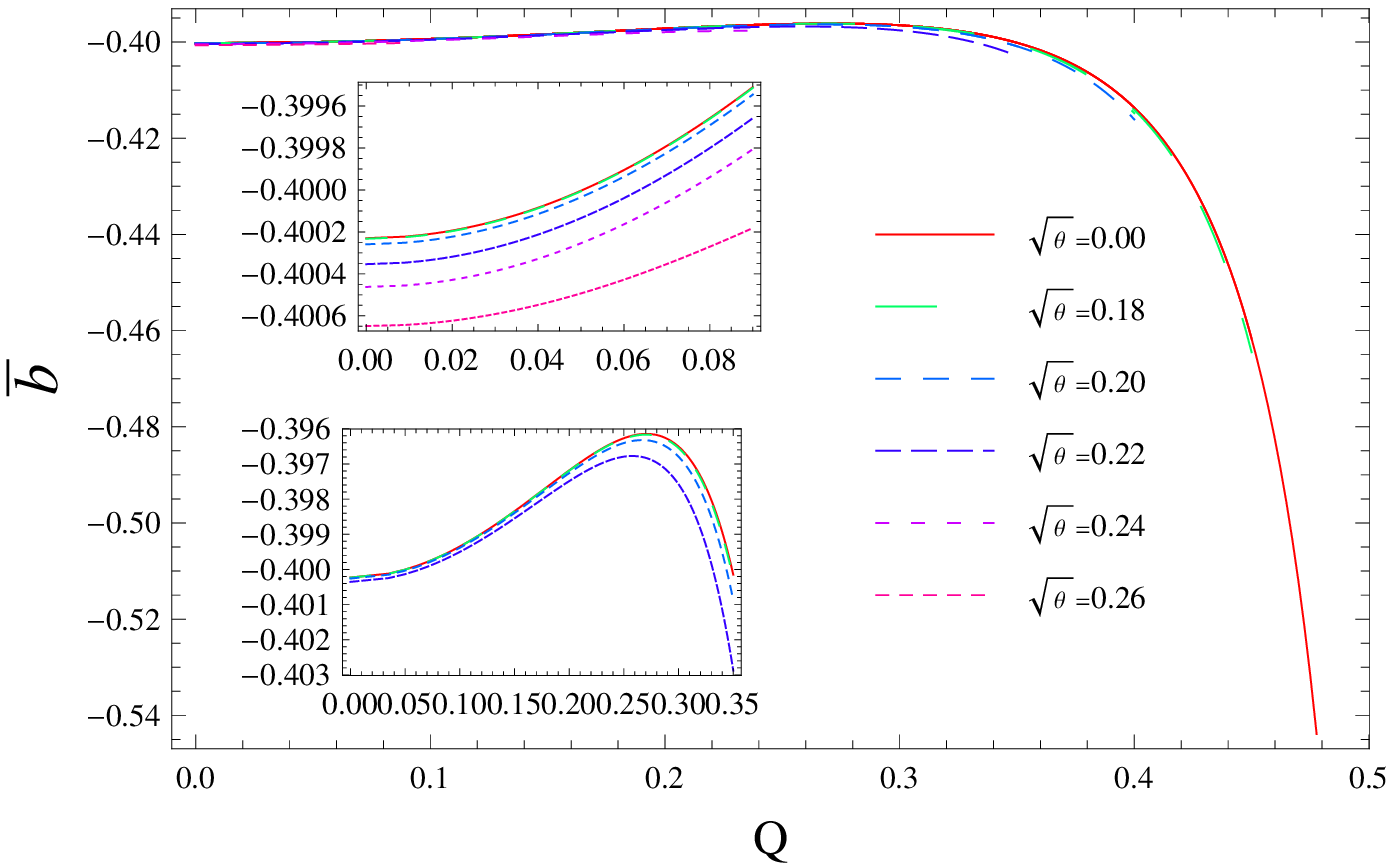}\;\;\;\;\;\\
\caption{Variation of the coefficients of the strong field limit
$\bar{a}$, $\bar{b}$ and the minimum impact parameter $u_{ps}$ with
the spacetime noncommutative parameter $\sqrt{\vartheta}$ in the
noncommutative Reissner-Nordstr\"{o}m black hole spacetime.
}\label{f1}
 \end{center}
 \end{figure}

Following the method developed by Bozza \cite{Bozza2,chen80}, we
define a variable
\begin{eqnarray}
z=1-\frac{r_0}{r},
\end{eqnarray}
and obtain
\begin{eqnarray}
I(r_0)=\int^{1}_{0}R(z,r_0)f(z,r_0)dz,\label{in1}
\end{eqnarray}
where
\begin{eqnarray}\label{3.8}
R(z,r_0)&=&\frac{2r_0\sqrt{A(r)B(r)C(r_0)}}{C(r)(1-z)^2}=2,
\end{eqnarray}
and
\begin{eqnarray}
f(z,r_0)&=&\frac{1}{\sqrt{A(r_0)-A(r)C(r_0)/C(r)}}.
\end{eqnarray}
The function $R(z, r_0)$ is regular for all values of $z$ and $r_0$.
However, $f(z, r_0)$ diverges as $z$ tends to zero. Thus, we split
the integral (\ref{in1}) into two parts
\begin{eqnarray}
I_D(r_0)&=&\int^{1}_{0}R(0,r_{ps})f_0(z,r_0)dz, \nonumber\\
I_R(r_0)&=&\int^{1}_{0}[R(z,r_0)f(z,r_0)-R(0,r_{ps})f_0(z,r_0)]dz
\label{intbr},
\end{eqnarray}
where $I_D(r_0)$ and $I_R(r_0)$ denote the divergent and regular
parts in the integral (\ref{in1}), respectively. To find the order
of divergence of the integrand, we expand the argument of the square
root in $f(z,r_0)$ to the second order in $z$ and obtain the
function $f_0(z,r_0)$:
\begin{eqnarray}
f_0(z,r_0)=\frac{1}{\sqrt{p(r_0)z+q(r_0)z^2}},
\end{eqnarray}
where
\begin{eqnarray}\label{3.12}
p(r_0)&=&2-\frac{6}{\sqrt{\pi}r_0}\gamma(\frac{3}{2},
\frac{r_{0}^2}{4\vartheta})+\frac{r_{0}^2}{2\vartheta\sqrt{\pi\vartheta}}
e^{-\frac{r_{0}^2}{4\vartheta}}+\frac{4Q^2}{\pi
r^2_{0}}F(r_{0})-\frac{2Q^2}{\pi r_0\sqrt{\vartheta}}G(r_{0}),\;
 \nonumber\\
q(r_0)&=&-1+\frac{6}{\sqrt{\pi}r_0}\gamma(\frac{3}{2},
\frac{r_{0}^2}{4\vartheta})-\frac{r_0^2}{4\vartheta\sqrt{\pi\vartheta}}e^{-\frac{r_0^2}{4\vartheta}}
\Big(2+\frac{r_0^2}{2\vartheta}\Big)-\frac{6Q^2}{\pi
r^2_{0}}F(r_{0}) \nonumber\\
&&+\frac{6Q^2}{\pi r_0\sqrt{\vartheta}}G(r_{0})-\frac{Q^2r^2_0}{2\pi
\vartheta^2}H(r_0),\nonumber\\
H(r_0)&=&\left[-\frac{\sqrt{\vartheta}}{r_0}\gamma(\frac{1}{2},
\frac{r_{0}^2}{4\vartheta})+e^{-\frac{r_0^2}{4\vartheta}}+\sqrt{2}-
\frac{r_{0}^2}{4\sqrt{2}\vartheta}\right]e^{-\frac{r_0^2}{4\vartheta}}.
\end{eqnarray}
When $r_0$ is equal to the radius of photon sphere $r_{ps}$, the
coefficient $p(r_0)$ vanishes and the leading term of the divergence
in $f_0(z,r_0)$ is $z^{-1}$; thus, the integral (\ref{in1}) diverges
logarithmically. Close to the divergence, the deflection angle can
be expanded in the form
\begin{eqnarray}
\alpha(\theta)=-\bar{a}\log{\bigg(\frac{\theta
D_{OL}}{u_{ps}}-1\bigg)}+\bar{b}+O(u-u_{ps}),
\end{eqnarray}
where
\begin{eqnarray}
\bar{a}&=&\frac{R(0,r_{ps})}{2\sqrt{q(r_{ps})}}=\frac{1}{\sqrt{q(r_{ps})}},\nonumber\\
q(r_{ps})&=&1-\frac{r_{ps}^4}{8\vartheta^2\sqrt{\pi\vartheta}
}e^{-\frac{r_{ps}^2}{4\vartheta}}-\frac{2Q^2}{\pi
r_{ps}^2}F(r_{ps})\nonumber\\
&&-\frac{Q^2r^2_{ps}}{2\pi \vartheta^2}H(r_{ps})+ \frac{4Q^2}{\pi
r_{ps}\sqrt{\vartheta}}G(r_{ps}), \nonumber\\
\bar{b}&=&
-\pi+b_R+\bar{a}\log{\frac{4q^2(r_{ps})\big[2A(r_{ps})-r_{ps}^2A''(r_{ps})\big]}{
p^{'2}(r_{ps})u_{ps}r_{ps}\sqrt{A^3(r_{ps})}}}, \nonumber\\
b_R&=&I_R(r_{ps}),\;\;\;\;\;p'(r_{ps})=\frac{dp}{dr_0}\big|_{r_0=r_{ps}},\;\;\;\;\;u_{ps}
=\frac{r_{ps}}{\sqrt{A(r_{ps})}}~.
\end{eqnarray}
$D_{OL}$ denotes the distance between the observer and the
gravitational lens, $\bar{a}$ and $\bar{b}$ are the so-called strong
field limit coefficients which depend on the metric functions
evaluated at $r_{ps}$ and, $u_{ps}$ is the minimum impact parameter.
Now we turn to determine $b_R$. From Eqs. (\ref{3.8}---\ref{3.12}),
we obtain
\begin{eqnarray}
&&b_R=I_R(r_{ps})=2\int^{1}_{0}g(z,\sqrt{\vartheta},Q)dz,
\;r_{ps}=r_{ps}(\sqrt{\vartheta},Q),\;r=r(z,\sqrt{\vartheta},Q),\nonumber\\
&&g(z,\sqrt{\vartheta},Q)=\frac{1}{\sqrt{A(r_{ps})-A(r)C(r_{ps})/C(r)}}-\frac{1}{\sqrt{q(r_{ps})}z}.
\label{}
\end{eqnarray}
For a certain charge $Q=Q_0$,
$b_R=2\int^{1}_{0}g(z,\sqrt{\vartheta})dz$. In general, the
coefficient $b_R$ can not be calculated analytically and, in this
case it cannot be evaluated numerically. Here we expand it in powers
of $\sqrt{\vartheta}$ as
\begin{eqnarray}
&&b_R=I_{R,0}+I_{R,1}\sqrt{\vartheta}+\frac{1}{2}I_{R,2}\sqrt{\vartheta}^2+\cdots,\nonumber\\
&&I_{R,0}=2\int^{1}_{0}g(z,\sqrt{\vartheta})|_{\sqrt{\vartheta}\rightarrow0}dz,\;
I_{R,1}=2\int^{1}_{0}\frac{dg}{d\sqrt{\vartheta}}|_{\sqrt{\vartheta}\rightarrow0}dz,\;\cdots.
\end{eqnarray}
 Because the values of various low
derivative of $g(z,\sqrt{\vartheta})$ at $\vartheta\rightarrow0$ is
zero, we can get
\begin{eqnarray}
b_R=I_{R,0}+ O(\sqrt{\vartheta}).
\end{eqnarray}
Then, we can obtain the $\bar{a}$, $\bar{b}$ and $u_{ps}$, and
describe them in Fig. \ref{f1}.

  Figure \ref{f1} tells us that given a certain charge, with
the increases of $\sqrt{\vartheta}$ the coefficient $\bar{a}$
increases, the $\bar{b}$ and the minimum impact parameter $u_{ps}$
decrease. It also shows that: i) when $\sqrt{\vartheta}$ is strong,
$Q_{max}$ is small, so that the coefficients $\bar{a}$, $\bar{b}$
and $u_{ps}$ are close to those of the noncommutative Schwarzschild
black hole; ii) when $\sqrt{\vartheta}$ is weak, $Q_{max}$ is big, so that the coefficients $\bar{a}$,
$\bar{b}$ and $u_{ps}$ are close to those of the commutative
Reissner-Nordstr\"{o}m black hole; iii) when $\sqrt{\vartheta}$ is
weak, the effect of spacetime noncommutativity is obvious till that
the charge quantity is very big; iv) the influence of the parameter
$\sqrt{\vartheta}$ on $\bar{a}$ and $\bar{b}$ is bigger than that on
$u_{ps}$. In principle we can distinguish a noncommutative
Reissner-Nordstr\"{o}m black hole from the commutative one and the
noncommutative Schwarzschild black hole, and then maybe probe the
value of the spacetime noncommutative constant by using strong field
gravitational lensing.

\begin{figure}[ht]
\begin{center}
\includegraphics[width=8.0cm]{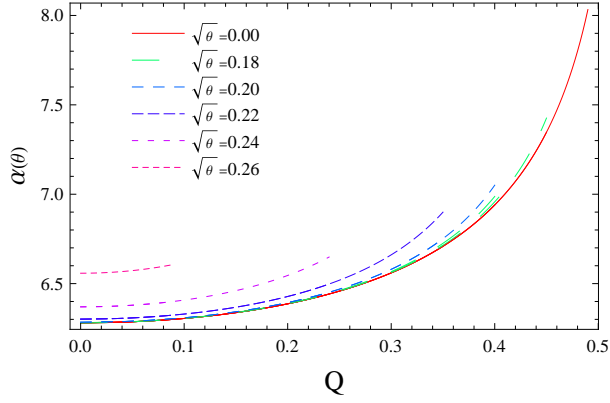}\;\;\;\;\;
\caption{ Deflection angles in the noncommutative
Reissner-Nordstr\"{o}m black hole spacetime evaluated at
$u=u_{ps}+0.00326$ as functions of $Q$.} \label{f2}
 \end{center}
 \end{figure}

Figure \ref{f2} shows the deflection angle $\alpha (\theta)$
evaluated at $u=u_{ps}+0.00326$. At this particular value of the
impact parameter $\alpha (\theta)=2\pi$ when
$Q=0,\sqrt{\vartheta}\rightarrow0$, which shows the photon just
takes one loop around the black hole and the relativistic images
begin to appear. Figure \ref{f2} indicates that the presence of
$\sqrt{\vartheta}$ increases the deflection angle $\alpha (\theta)$
for the light propagated in the noncommutative
Reissner-Nordstr\"{o}m black hole spacetime. Comparing with those in
the commutative one, we could extract the information about the size
of spacetime noncommutative parameter $\sqrt{\vartheta}$ by using
strong field gravitational lensing.

Considering the source, lens and observer are highly aligned, the
lens equation in strong gravitational lensing can be written as
\cite{Bozza1}
\begin{eqnarray}
\beta=\theta-\frac{D_{LS}}{D_{OS}}\Delta\alpha_{n},
\end{eqnarray}
where $D_{LS}$ is the distance between the lens and the source,
$D_{OS}=D_{LS}+D_{OL}$, $\beta$ is the angular separation between
the source and the lens, $\theta$ is the angular separation between
the image and the lens, $\Delta\alpha_{n}=\alpha-2n\pi$ is the
offset of deflection angle and $n$ is an integer. The position of
the $n$-th relativistic image can be approximated as
\begin{eqnarray}
\theta_n=\theta^0_n+\frac{u_{ps}e_n(\beta-\theta^0_n)D_{OS}}{\bar{a}D_{LS}D_{OL}},
\end{eqnarray}
where
\begin{eqnarray}
e_n=e^{\frac{\bar{b}-2n\pi}{\bar{a}}},
\end{eqnarray}
$\theta^0_n$ are the image positions corresponding to
$\alpha=2n\pi$.  The magnification of the $n$-th relativistic image
is given by
\begin{eqnarray}
\mu_n=\frac{u^2_{ps}e_n(1+e_n)D_{OS}}{\bar{a}\beta D_{LS}D^2_{OL}}.
\end{eqnarray}
If $\theta_{\infty}$ represents the asymptotic position of a set of
images in the limit $n\rightarrow \infty$, the minimum impact
parameter $u_{ps}$ can be simply obtained as
\begin{eqnarray}
u_{ps}=D_{OL}\theta_{\infty}.
\end{eqnarray}
In the simplest situation, we consider only that the outermost image
$\theta_1$ is resolved as a single image and all the remaining ones
are packed together at $\theta_{\infty}$. Then the angular
separation between the first image and other ones can be expressed
as
\begin{eqnarray}
s=\theta_1-\theta_{\infty},
\end{eqnarray}
and the ratio of the flux from the first image and those from the
all other images is given by
\begin{eqnarray}
\mathcal{R}=\frac{\mu_1}{\sum^{\infty}_{n=2}\mu_{n}}.
\end{eqnarray}
For a highly aligned source, lens and observer geometry, these
observables can be simplified as
\begin{eqnarray}
&s&=\theta_{\infty}e^{\frac{\bar{b}-2\pi}{\bar{a}}},\nonumber\\
&\mathcal{R}&= e^{\frac{2\pi}{\bar{a}}}.
\end{eqnarray}
The strong deflection limit coefficients $\bar{a}$, $\bar{b}$, and
the minimum impact parameter $u_{ps}$ can be obtained through
measuring $s$, $\mathcal{R}$, and $\theta_{\infty}$. Then, comparing
their values with those predicted by the theoretical models, we can
identify the nature of the black hole lens.

\section{Numerical estimation of observational gravitational lensing parameters}

In this section, supposing that the gravitational field of the
supermassive black hole at the Galactic center of the Milky Way can
be described by the noncommutative Reissner-Nordstr\"{o}m black hole
metric, we estimate the numerical values for the coefficients and
observables of the strong gravitational lensing; then we study the
effect of the spacetime noncommutative parameter $\sqrt{\vartheta}$
on the gravitational lensing.

\begin{figure}[ht]
\begin{center}
\includegraphics[width=8.0cm]{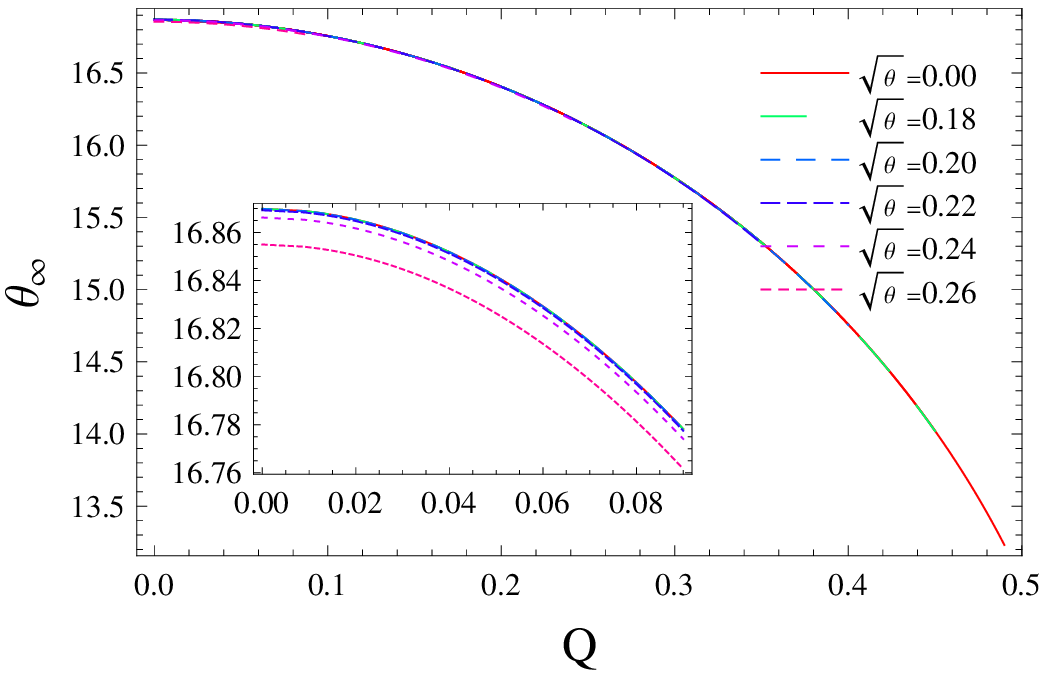}\;\;\;\;\;\\
\includegraphics[width=7.50cm]{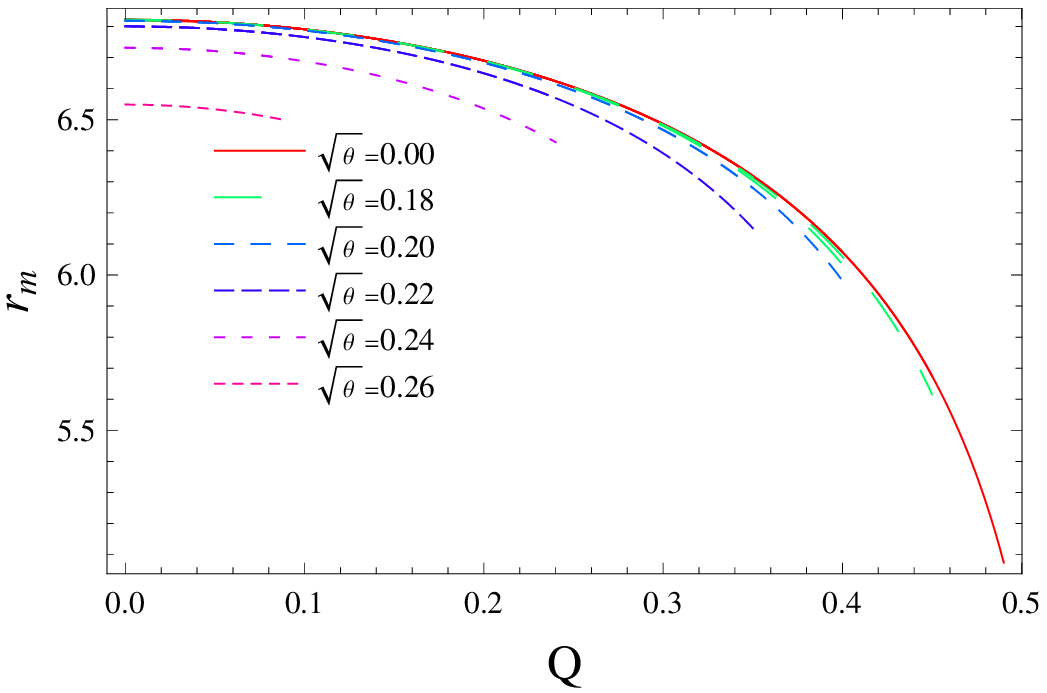}\;\;\;\;\;\includegraphics[width=7.50cm]{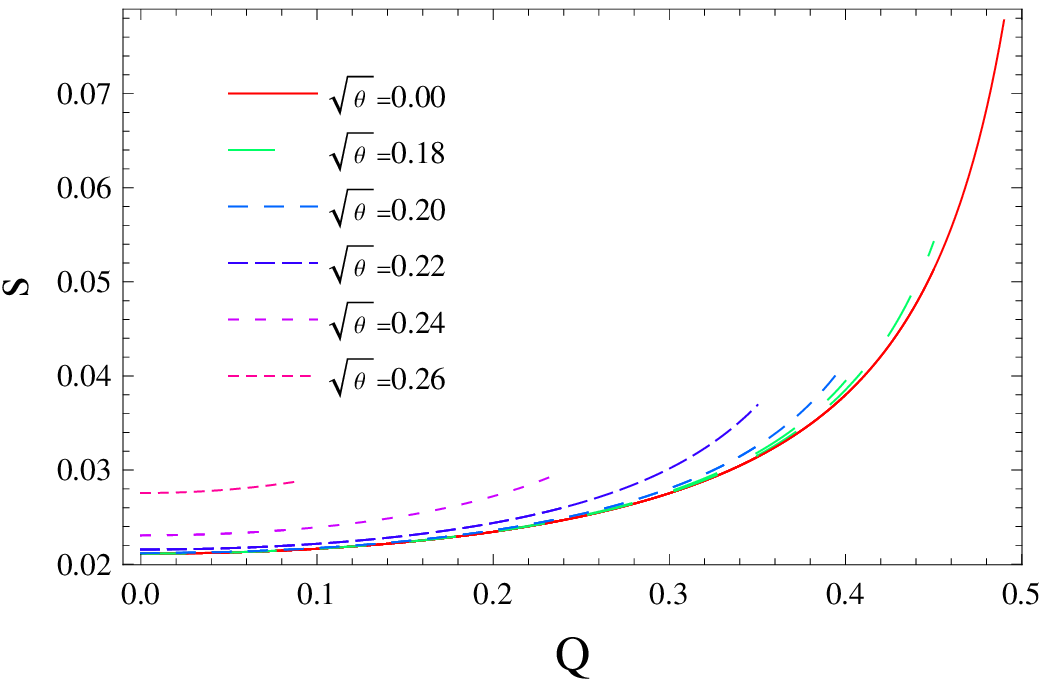}\\
\caption{Strong gravitational lensing by the Galactic center black
hole. Variation of the values of the angular position
$\theta_{\infty}$, the relative magnitudes $r_m$, and the angular
separation $s$ with parameter $Q$ in the noncommutative
Reissner-Nordstr\"{o}m black hole spacetime.} \label{f3}
 \end{center}
 \end{figure}

The mass of the central object of our Galaxy is estimated to be
$2.8\times 10^6M_{\odot}$ and its distance is around $8.5$ kpc
\cite{richs}. For different $\vartheta$, the numerical value of the
minimum impact parameter $u_{ps}$, the angular position of the
asymptotic relativistic images $\theta_{\infty}$, the angular
separation $s$, and the relative magnification of the outermost
relativistic image with the other relativistic images $r_{m}$ are
listed in Table \ref{table} and \ref{table3}, and described in Fig.
\ref{f3}.
\begin{table}[h]
\caption{Numerical estimation for main observables and the strong
field limit coefficients for the black hole at the center of our
Galaxy, which is supposed to be described by the noncommutative
Reissner-Nordstr\"{o}m black hole with the charge $Q=0.09$. $R_s$ is
Schwarzschild radius. $r_m=2.5\log{\mathcal{R}}$.}\label{table}
\begin{center}
\begin{tabular}{|c|c|c|c|c|c|c|}
\hline \hline $\sqrt{\vartheta}$ &$\theta_{\infty}
$($\mu$arcsecs)&\; $s$ ($\mu$arcsecs) \;\; & $r_m$(magnitudes)
&\;\;\;\;$u_{ps}/R_S$\;\;\;\; &
\;\;\;\;\;\;\;\;$\bar{a}$\;\;\;\;\;\;\;\; &\;\;\;\;\;\;\;\;
$\bar{b}$\;\;\;\;\;\;\;\; \\
\hline
 0& 16.7782& 0.02153& 6.79692&2.58396& 1.00367& $-0.399510$ \\
 \hline
0.16&16.7782&0.02153&6.79691&2.58396& 1.00367&$-0.399510$\\
\hline
0.18& 16.7782&0.02154&6.79669&2.58396&1.00371&$-0.399514$ \\
\hline
0.20& 16.7781&0.02161&6.79342&2.58395&1.00419&$-0.399546$ \\
 \hline
0.22 &16.7775&0.02205&6.77263&2.58385&1.00727& $-0.399658$\\
 \hline
0.24&16.7741&0.02375&6.69627&2.58332&1.01876&$-0.399806$\\
 \hline
0.26&16.7616&0.02883&6.49748&2.58140&1.04993&$-0.400182$
 \\
\hline\hline
\end{tabular}
\end{center}
\end{table}
\begin{table}[h]
\caption{Numerical estimation for main observables and the strong
field limit coefficients for the black hole at the center of our
Galaxy, which is supposed to be described by the noncommutative
Reissner-Nordstr\"{o}m black hole with the charge $Q=0.49$. $R_s$ is
Schwarzschild radius. $r_m=2.5\log{\mathcal{R}}$.}\label{table3}
\begin{center}
\begin{tabular}{|c|c|c|c|c|c|c|}
\hline \hline $\sqrt{\vartheta}$ &$\theta_{\infty}
$($\mu$arcsecs)&\; $s$ ($\mu$arcsecs) \;\; & $r_m$(magnitudes)
&\;\;\;\;$u_{ps}/R_S$\;\;\;\; &
\;\;\;\;\;\;\;\;$\bar{a}$\;\;\;\;\;\;\;\; &\;\;\;\;\;\;\;\;
$\bar{b}$\;\;\;\;\;\;\;\; \\
\hline
 0& 13.2321& 0.0777923& 5.07464&2.03784& 1.34431& $-0.621671$ \\
 \hline
0.09&13.2321&0.0777923&5.07464&2.03784& 1.34431&$-0.621671$\\
\hline
0.10& 13.2321&0.0777923&5.07464&2.03784&1.34431&$-0.621671$ \\
\hline
0.11& 13.2321&0.0777923&5.07464&2.03784&1.34431&$-0.621671$ \\
 \hline
0.12 &13.2321&0.0777939&5.07461&2.03784&1.34432& $-0.621673$\\
 \hline
0.13&13.2321&0.0778157&5.07431&2.03783&1.34440&$-0.621707$\\
 \hline
0.14&13.2321&0.0779688&5.07220&2.03783&1.34495&$-0.621932$
 \\
\hline\hline
\end{tabular}
\end{center}
\end{table}

From Table \ref{table} and \ref{table3}, it is easy to see that our
results reduce those in the commutative Reissner-Nordstr\"{o}m black
hole sacetime \cite{Bozza2} to $\sqrt{\vartheta}\rightarrow0$. Table
\ref{table} shows that if $\sqrt{\vartheta}$ is strong, for example,
$\sqrt{\vartheta}=0.26$, the observables $\theta_\infty$, $r_m$ and
$s$ are close to those of the noncommutative Schwarzschild black
hole \cite{ding}. Moreover, we also find that as the parameter
$\sqrt{\vartheta}$ increases, the minimum impact parameter $u_{ps}$,
the angular position of the relativistic images $\theta_{\infty}$,
and the relative magnitudes $r_m$ decrease, but the angular
separation $s$ increases. From Fig. \ref{f3}, we find that the
influence of the parameter $\sqrt{\vartheta}$ on $r_m$ and $s$ is
bigger than that on $\theta_\infty$. This means that the bending
angle is bigger and the relative magnification of the outermost
relativistic image with the other relativistic images is smaller in
the noncommutative Reissner-Nordstr\"{o}m black hole spacetime.

From Tab. \ref{table3} and Fig. \ref{f3} we also find that if
$\sqrt{\vartheta}$ is weak, e.g. $0.12\leq\sqrt{\vartheta}\leq0.14$,
the effect of spacetime noncommutativity is obvious till that the
charge quantity is very big, e.g., $Q=0.49$. Tab. \ref{table3} also
shows that the detectable lower limit of $\sqrt{\vartheta}$ is
$0.12$. From Tab. \ref{tabel1}, we can let the charge even bigger,
i.e. $0.49<Q<0.5$, then the detectable lower limit lies between
$0.11$ and $0.12$. But it is only a numerical result, so the
detectable lower limit $0.12$ is more appropriate. Therefore we can
conclude that the detectable scope of $\vartheta$ in a
noncommutative Reissner-Nordstr\"{o}m black hole lensing is
$0.12\leq\sqrt{\vartheta}\leq0.26$.

In order to identify the nature of these three compact objects
lensing, it is necessary for us to measure angular separation $s$
and the relative magnification $r_m$ in the astronomical
observations. Table \ref{table3} tells us that the resolution of the
extremely faint image is $\sim 0.07$ $\mu$ arc sec, which is too
small. However, with the development of technology, the effects of
the spacetime noncommutative constant $\sqrt{\vartheta}$ on
gravitational lensing may be detected in the future.

\section{Summary}

Spacetime noncommutative constant would be a new fundamental natural
constant which can affect the classical gravitational effect such as
gravitational lensing.
 Studying the strong gravitational lensing can help
us to probe the spacetime noncommutative constant. In this paper we
have investigated strong field lensing in the noncommutative
Reissner-Nordstr\"{o}m black hole spacetime to study the influence
of the spacetime noncommutative parameter on the strong
gravitational lensing. The model was applied to the supermassive
black hole in the Galactic center. Our results show that with the
increase of the parameter $\sqrt{\vartheta}$, the minimum impact
parameter $u_{ps}$, the angular position of the relativistic images
$\theta_{\infty}$ and the relative magnitudes $r_m$ decrease, and
the angular separation $s$ increases.

Our results also show that i) if $\sqrt{\vartheta}$ is strong,
$Q_{max}$ is small, so that the observables $\theta_\infty$, $r_m$
and $s$ are close to those of the noncommutative Schwarzschild black
hole lensing; ii) if $\sqrt{\vartheta}$ is weak, $Q_{max}$ is big,
so that the observables $\theta_\infty$, $r_m$ and $s$ are close to
those of the commutative Reissner-Nordstr\"{o}m black hole lensing;
iii) if $\sqrt{\vartheta}$ is weak, e.g.
$0.12\leq\sqrt{\vartheta}\leq0.14$, the effect of spacetime
noncommutativity is obvious till that the charge quantity is very
big, and if $\sqrt{\vartheta}<0.12$, we cannot probe the $\vartheta$
at all; iv) the influence of the parameter $\sqrt{\vartheta}$ on
$r_m$ and $s$ is bigger than that on $\theta_\infty$.

In short, the detectable scope of $\vartheta$ in a noncommutative
Reissner-Nordstr\"{o}m black hole lensing is
$0.12\leq\sqrt{\vartheta}\leq0.26$, which is wider than that in a
noncommutative Schwarzschild black hole lensing,
$0.18\leq\sqrt{\vartheta}\leq0.26$ \cite{ding}.
 This may offer a way
to distinguish a noncommutative Reissner-Nordstr\"{o}m black hole
from the commutative one and the noncommutative Schwarzschild black
hole and probe the spacetime noncommutative constant with the
astronomical instruments in the future.

\begin{acknowledgments}
This work was partially supported by Hunan Provincial Natural
Science Foundation of China under Grant No. 11JJ3014 and the
Scientific Research Foundation for the introduced talents of Hunan
Institute of Humanities, Science and Technology. J. Jing's work was
supported by the NNSFC No.10875040 and No.10935013, 973 Program No.
2010CB833004 and PCSIRT under Grant No. IRT0964.
\end{acknowledgments}


\begin{thebibliography}{99}
\bibitem{foot} The notation $\vartheta$ used here is a constant as well as Plank
constant $\hbar$, but we still call it a spacetime noncommutative
parameter since it up to now is undetermined.
\end{thebibliography}
\end{document}